# Finding and Solving Contradictions of False Positives in Virus Scanning

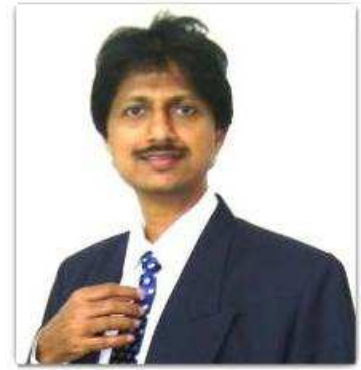

**By- Umakant Mishra, Bangalore, India**
umakant@trizsite.tk, http://umakant.trizsite.tk

**Contents**

## 1. What is false Positive

False positive is falsely and positively identifying a virus, i.e., an instance of wrongly labeling a benign program as malicious program. False positive is considered as a drawback of a virus detection method. Minor weaknesses of any virus detection method can lead to false positives. But ideally the virus detection method should not lead to any false positive. False positives are equally dangerous as false negatives. The false positive rate is calculated in the following formula.

$$\text{False Positive Rate} = \frac{\text{total number of false positive instances}}{\text{total number of benign instances}}$$

Ideally the false positive rate should remain 0 or very close to 0. Even a slightest increase in false positive rate is considered as undesirable.

## 2. Probability of False Positives in different Methods of Virus Scanning

Some methods of virus scanning are more susceptible to false positives than others. Before discussing more about false positives it may be useful to give a brief on different methods of virus scanning.

**Specific signature scanning**

Most computer viruses copy themselves to each file they infect. They replicate an identical copy of themselves byte-by-byte each time they infect a new file. These types of viruses can be detected by searching virus signatures. This method is more reliable and rarely leads to any false positives.

**Generic Signature Scanning**

Generic signature uses a signature pattern that is found in a family of viruses. The generic signatures use various wildcards to detect all the variants of a virus family. This is a quicker method and capable of detecting new and future viruses of the same family but may lead to false positives.

**Integrity checking**

An Integrity checker detects the existence of viruses by comparing the hash values of a file with the hash value of its uninfected version. If no difference is found between the two hash values then the file is deemed to be uninfected. This

Eliminating False Positives in Virus Scanning, by Umakant Mishra				http://www.trizsite.tk

method works because the viruses must make changes to their host programs in order to infect.

**Behavior monitoring**

The method of behavior monitoring tries to detect virus type activities, such as, attempting to format a disk, or moving a file into the operating system folders etc. which is generally not done by a common program. These actions are flagged as questionable or suspicious by behavior monitors.

**Heuristic scanning**

A heuristic engine detects the commands within a program that are not found in typical application programs, such as, the replication mechanism of a virus, the distribution routine of a worm or the payload of a Trojan. If the target program's code appears to be virus-like, then scanner reports a possible infection.

**Network anomaly detection**

This method checks the anomaly of data transfers within a network to detect a virus infection. It may also look for certain codes or signatures in network packets. If the packets are found to be suspicious then an alarm is raised.

**Mixed heuristic detection**

Nowadays all anti-viruses follow a combination of various heuristic detection methods along with generic detection methods in order to authenticate detection and avoid false positives. A mixed heuristic may include static and dynamic heuristics and other methods.

| Methods of virus detection | Chances of false positives |
|---|---|
| Integrity Checking | NIL |
| Specific signature scanning | NIL – extremely low |
| Generic signature scanning | Low – Medium |
| Behavior monitoring | Medium – High |
| Heuristic detection | Medium – High |
| Network anomaly detection | Medium- High |
| Other Generic methods | Medium- High |
| A mixture of multiple methods | Extremely low |

## 3. Reasons of getting false positives

The specific methods provide very accurate scanning by comparing viruses with their exact signatures. But these methods fail to detect new and unknown viruses. On the other hand the generic methods can detect even new viruses without using virus signatures. But these methods are more likely to generate false positives. There is a positive correlation between the capability to detect new and unknown viruses and false positive rate.

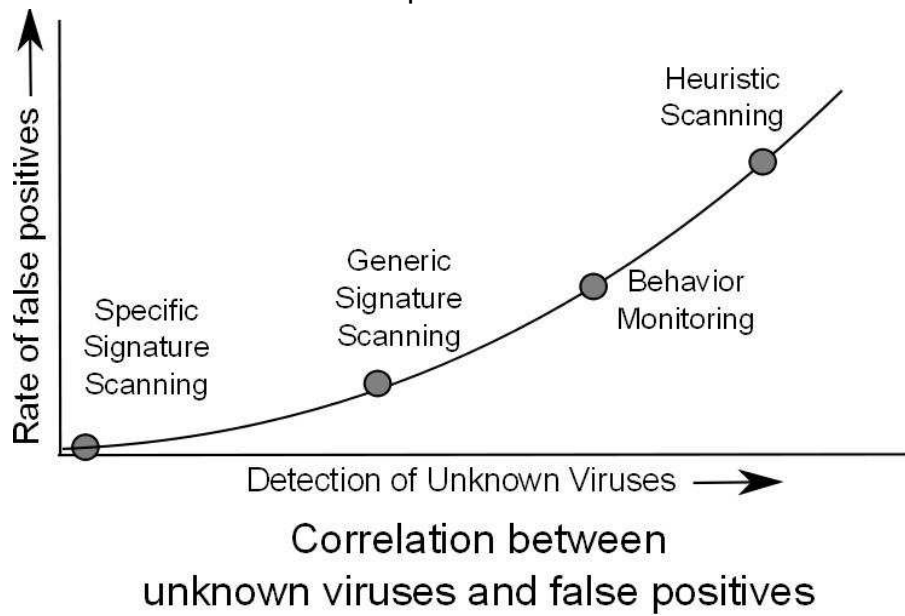

Correlation between unknown viruses and false positives

For example, the heuristic methods can detect new and unknown viruses but they are more susceptible to false positives. This is because the inherent strategy followed by the heuristics are based on probabilistic methods and do not guarantee an infection. For example, if a static heuristic antivirus program finds a "file open" operation, followed by "file read" and "write" operations, and also finds a string "VIRUS" in the program, then it may report that the file is infected by an unknown virus. But there is a possibility that a genuine file satisfies all these conditions.

Thus whenever a detection method attempts to detect a new and unknown virus the attempt may result in false positive. Therefore generic methods are more susceptible to false positives. The following are some of the reasons of getting false positive alerts.

In some cases it is extremely difficult to differentiate between the identity and behavior of good and bad code. Any mistake in judgment can lead to false positives or false negatives.

An anti-virus program searches for only signatures and not the complete virus code. Besides in many cases it searches for wild card signatures. These situations lead to many matches that are not true positives.

As the conventional signature scanning does not work for polymorphic and metamorphic viruses, the advanced anti-virus programs have to use various heuristic methods to detect such viruses. But such methods raise a high rate of false positives.

Certain situations like stream scanning of video files requires very fast scan, where the scanner may have to compromise some aspects of reliability. Besides in a stream scanning, only the packets are available to the scanner. Hence a file level scanning is not possible.

## 4. The Ideal Final Result and Contradictions

The TRIZ concept of Ideal Final result suggests us to determine what is ideally expected in a problem situation irrespective of whether it is possible or not. The ideality of a product or process or situation is achieved when all the functions of the product or process are achieved without facing any problems or cost. Ideality is the sum of all benefits divided by all costs. If we consider an ideal situation of eliminating false positives we find the following IFR.

> **IFR** **Ideally the anti-virus should detect and prevent viruses with full certainty. The chances of error should be nil and it should not raise any false positive or false negative.**

While a traditional approach tries to achieve a right balance between false positives and false negatives a TRIZ approach looks forward to achieve the above IFR of avoiding all sorts of false positives and false negatives. The TRIZ approach first tries to find the contradictions and then solve the contradictions. Each problem related to false positives is presented in the form of a contradiction. Once a contradiction is formulated the problem becomes very clear and solvable by using various TRIZ techniques. The following are some of the contradictions related to false positives in virus scanning.

## 5. Contradiction of tighter rules and false positives

It is very difficult to avoid both false positives and false negatives. If we tight the set of rules we will fail to detect many viruses. If we loosen the set of rules we will get more false positives.

> **Contradiction:** Virus detection involves various sets of rules, whether signatures scanning or heuristic scanning. If we have tighter set of rules, then we may miss many real viruses. On the other hand, if we have looser set of rules we will detect many false positives. We want to detect viruses and not false positives.

**Possible solutions**

- One option is to use slightly tighter rules but multiple rules and use multiple detection methods. The tighter rules will avoid false positives. Although the method will skip some genuine viruses the viruses cannot ultimately escape as they will be detected by other rules and detection methods.

- Patent 8239948 suggests to use a computer implemented method to find multiple signatures of a virus and score all these signatures according to their likelihood to be found in non-malicious programs. Finally the candidate signature is selected based on such scores. As the signatures are pre tested to be unique, the chances of false positives are reduced.

## 6. Contradiction of confirming the infection

Generally an anti-virus program generates an alarm signal when there is a reasonable suspicion. Waiting for more time for a confirmed detection will allow the virus to do its intended damages.

> **Contradiction:** If the anti-virus waits longer to get sufficient proof about a virus operation then the virus may make undesirable changes and damages to the program. On the other hand, if the anti-virus suspects an operation before it causes any infection then the detection can be a false positive.

**Possible solutions**

- One solution is to use emulation techniques to run the virus in a virtual machine. This method allows the virus to infect the virtual machine but the actual machine remains safe and uninfected (Principle-26: Copy).

## 7. Contradiction of raising virus alarm

Sometimes the anti virus program finds that a system file has been modified but cannot determine whether the file has been modified by a virus or by the user. In such a situation, if the program generates a virus alarm it may lead to a false positive.

> **Contradiction**: If the anti-virus program is not definitive about a suspicious alternation in a system file and raises a virus alarm then it may lead to a false positive. On the other hand if it ignores such a suspicious alteration then it may allow a virus to cause serious damages. Both the situations are dangerous.

**Possible Solutions**

- One solution to the above problem is that the anti-virus would ask the user to take a decision (Principle-23: Feedback). If the user knows the operation to be genuine and agrees to ignore it then anti-virus ignores the virus like operation. **Drawback**: the user may not be adequately knowledgeable to take such a decision.

## 8. Contradiction of adequate testing of anti-virus software

Every product must go through a rigorous testing process before being released to the market. But there is a problem in testing anti-virus programs satisfactorily, as there are frequent updates in virus signatures and detection algorithms.

> **Contradiction**: If the vendor performs more comprehensive internal testing of the software update then there will be a delay in releasing the product. On the other hand if the vender releases the software updates without delay, then the testing remains insufficient and involves the risk of false positives. The vender wants to release the updates fast but doesn't want to take the risk of false positives.

**Possible Solutions:**

- The anti-virus companies may adopt computerized techniques to fast extract signatures and automated methods for fast testing the software (Principle-21: Skip, Principle-25: Self Service).

- Another option is to use an online scanning service provided by an anti-virus company and scan the computer remotely (Principle-17: Another dimension). This method allows the client's files to be scanned with the latest signatures.

- Patent 7290282 (invented by Renert, et al., Assignee- Symantec Corporation. Oct. 2007, also Patent 7757292, by the same inventors, Jul 2010) discloses to release the virus detection technique with a constraints module that which allows specific clients to test the new heuristics and provide test reports to the software server (Principle-23: Feedback). If the server administrator (or vender) finds more false positives in the feedback then he modifies the constraints and re-release to the clients (Principle-35: Parameter change). Once the virus detection technique is ready for wide release, the technique is circulated to all the clients.

## 9. Contradiction of too many signatures and false positives

The increase in the virus population results in an increase in the number of virus signatures. But a large number of virus signatures lead to various disadvantages.

> **Contradiction**: If the signature database contains all virus signatures then it becomes big and difficult to download. If the size of the signature database is maintained small then it can contain less number of signatures and will generate more false negatives (or false positives). The size of the database should be small to download but should be large to contain all virus signatures.

**Possible Solutions**

- The signature database is compressed before downloading (Principle-37: Expansion and Compression)

- Using second and third generation heuristic scanning which is quite sophisticated. These heuristics have high capabilities of detecting stealth viruses and also capabilities of detecting with high accuracy (Principle-28: Mechanics Substitution).

## 10. Contradiction of scanning time and false positives

A slow scanning is irritating to most users. But a faster scanning may not be sufficiently reliable and may lead to more false positives or false negatives. In most cases there is a tradeoff between speed and thoroughness.

> **Contradiction:** If a thorough scanning is done using all signatures and application of all methods on all the files then the scanning will take enormous time and may not finish before the next scanning is due. On the other hand, if some signatures will be ignored or some methods will be omitted then the scanning may not be reliable and may lead to false positives. We want high speed scanning but don't want false positives.

**Possible Solutions**

- The traditional approach uses single, specific signature types to detect viruses, i.e., one virus - one signature. This method raises more false positives. Patent 6338141 proposes RAVEN detection system that uses a combination of multiple signatures and flags (up to about 70 different data items depending on the virus type). Using multiple signatures for each virus allows RAVEN to verify infections with a high degree of certainty and avoid possibility of false identifications.

## 11. Contradiction of false positives in heuristic scanning

The heuristic scanners use a threshold level to determine whether a code is malicious or not. If the heuristic score is above the threshold level then the program is detected as suspicious. Typically a medium level threshold is used to give a balance between speed and accuracy.

> **Contradiction:** If the threshold level is higher then many suspicious activities will be ignored. On the other hand if the threshold level is lower then it increases false positives. We want to increase the threshold to avoid false positives and decrease the threshold to avoid false negatives.

**Possible Solutions**

- The heuristic scanners generally consider a large number of heuristics each heuristics having a specific weightage. If the summed weightage exceeds a threshold then the program is deemed to be infected. As the weightage is calculated by summing up multiple heuristic events the possibility of false positives or false negatives is reduced.

- Patent 8028338 (Invented by Schneider et. al., Sep 2011) discloses a set of goodware characteristics which can be used to determine whether a file belongs to goodware. This method reduces detection of false positives.

- Another method is to integrate the heuristics of normal behaviors within the target program itself at the time of development. Patent 8195953 suggests to include an ability section in the programs which defines the capability of the program. If the actual behavior during execution varies from its internally defined capability then the program is suspected as infected. This method is different from the conventional method of defining the heuristics externally in the anti-virus program.

## 12. Throttling false positives in heuristic scanning

Heuristic scanning is very useful for detecting polymorphic, metamorphic and advanced complicated viruses. But heuristic scanning is based on various assumptions and can lead to false positives. There is a need to control false positives in heuristic scanning.

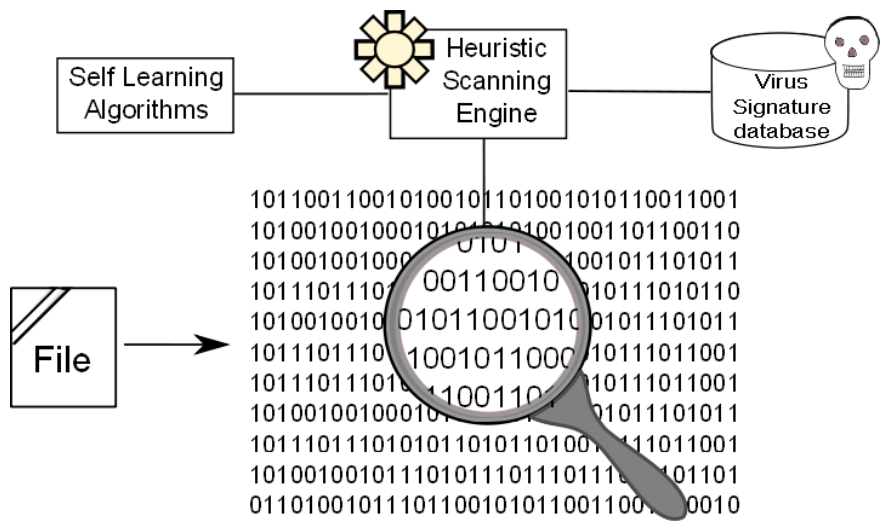

Heuristic Scanning by anti-virus programs

The heuristics of the second generation of scanners are much more sophisticated. While all the capabilities of the first generation scanners are retained many advanced heuristics are added, such as, code analysis, code tracing, strange opcodes etc. A modern heuristics includes artificial intelligence, self learning systems, behavior rules and even automatic creation of rules. Looking at various instances of the virus, the scanner learns which part of the virus can be altered and which part of the virus cannot. This learning of the intelligent scanner helps it to scan more and more variants of the same virus more confidently without leading to any false positives.

**Calculation of heuristics**

Heuristic scanning is based on calculation on each individual heuristics. Hence a single virus like characteristic is never determined as a virus. Availability of multiple heuristics creates a virus alarm. Besides, each virus like activity is assigned a weightage. After verifying all the behaviors the weightages are summed up to find whether the summed weightage exceeds a threshold. If the combined weightage exceeds the threshold then the program is deemed to be infected.

The method of heuristic calculation is similar whether the heuristics are used to determine file infection or system infection or email infection or server infection or network infection or others. For example in a network anomaly detection, several heuristics are considered each having different weights. The network packets are observed for a period that is neither too short nor too long.

> **Contradiction:** If the network packets are observed for a short period then the accuracy of observation will be low which will lead to false positives. If on the other hand the network packets will be observed for a long period then there is a chance of infection during the period of observation. The period should be short enough to avoid infection and should be long enough to ensure accuracy of observation.

The heuristic value is calculated by summing up the values of all the heuristics. But each heuristic does not have the same weightage. Some heuristics are more critical and therefore carry more weightage than others. Thus in a situation where there are multiple heuristics (H1, H2, … Hn) and each of them have different weightage (W1, W2, … Wn), the calculation of heuristic is the sum of all individual weightages of all the heuristic features.

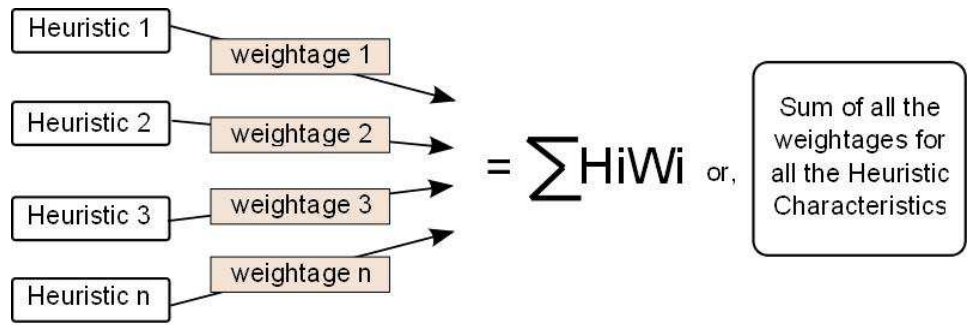

Calculation of Heuristic Value

**Adjusting threshold to control false positives**

However, in heuristic scanning, the threshold level of heuristic detection may be adjusted to improve the accuracy and speed of scanning.

> Highest level threshold- for exact identification or nearly exact identification. Minimum heuristics used.
>
> Medium level threshold- a balance between speed and accuracy. Speed scanning with low risk of false positives.
>
> Low-level threshold- the scanner becomes more sensitive to detect most of the new viruses/ malware. But the risk of false positives is increased.